\RequirePackage{lineno}
\documentclass[aps,twocolumn,showpacs,byrevtex,prl,reprint]{revtex4-1}

\usepackage{graphicx}% Include figure files
\usepackage{dcolumn}% Align table columns on decimal point
\usepackage{bm}% bold math
\usepackage{rotating}
\usepackage{epstopdf}
\usepackage{color}
\usepackage{verbatim} %for comment
\usepackage{multirow}
\usepackage[abs]{overpic}

\newcommand{\PreserveBackslash}[1]{\let\temp=\\#1\let\\=\temp}
\newcolumntype{C}[1]{>{\PreserveBackslash\centering}p{#1}}
\newcolumntype{R}[1]{>{\PreserveBackslash\raggedleft}p{#1}}
\newcolumntype{L}[1]{>{\PreserveBackslash\raggedright}p{#1}}

\newcommand{\EE}{e^+e^-}

\newcommand{\too}{\rightarrow}

\begin{document}
%\linenumbers
\graphicspath{{figure/}}
\DeclareGraphicsExtensions{.eps,.png,.ps}

\preprint{\vbox{\hbox{}
                \hbox{BESIII Analysis Draft - V1.1}
                }}

\title{\quad\\[0.0cm] \boldmath Observation of $h_{c}$ radiative decay $h_{c} \too \gamma \eta'$ and evidence for $h_{c} \too \gamma \eta$}

%\author{Author list here}
\author{
      M.~Ablikim$^{1}$, M.~N.~Achasov$^{9,e}$, X.~C.~Ai$^{1}$,
      O.~Albayrak$^{5}$, M.~Albrecht$^{4}$, D.~J.~Ambrose$^{44}$,
      A.~Amoroso$^{49A,49C}$, F.~F.~An$^{1}$, Q.~An$^{46,a}$,
      J.~Z.~Bai$^{1}$
% ...
      \\
      \vspace{0.2cm}
      (BESIII Collaboration)\\
      \vspace{0.2cm} {\it
        $^{1}$ Institute of High Energy Physics, Beijing 100049, People's Republic of China\\
% ...
$^{4}$ Bochum Ruhr-University, D-44780 Bochum, Germany\\
        $^{5}$ Carnegie Mellon University, Pittsburgh, Pennsylvania 15213, USA\\
% ...
$^{9}$ G.I. Budker Institute of Nuclear Physics SB RAS (BINP), Novosibirsk 630090, Russia\\
% ...
$^{44}$ University of Rochester, Rochester, New York 14627, USA\\
% ...
$^{46}$ University of Science and Technology of China, Hefei 230026, People's Republic of China\\
% ...
$^{49}$ (A)University of Turin, I-10125, Turin, Italy; (B)University of Eastern Piedmont, I-15121, Alessandria, Italy; (C)INFN, I-10125, Turin, Italy\\
% ...
$^{a}$ Also at State Key Laboratory of Particle Detection and Electronics, Beijing 100049, Hefei 230026, People's Republic of China\\
% ...
$^{e}$ Also at the Novosibirsk State University, Novosibirsk, 630090, Russia\\
% ...
      }
    \vspace{0.4cm}
}

%%% Local Variables:
%%% mode: latex
%%% TeX-master: "hc"
%%% End:

\begin{abstract}
A search for radiative decays of the $P$-wave spin singlet charmonium resonance $h_c$ is performed based on $4.48 \times 10^{8}$ $\psi'$ events collected with the BESIII detector operating at the BEPCII storage ring. Events of the reaction channels $h_{c} \too \gamma \eta'$ and $\gamma \eta$ are observed with a statistical significance of $8.4 \sigma$ and $4.0 \sigma$, respectively, for the first time. The branching fractions of $h_{c} \too \gamma \eta'$ and $h_{c} \too \gamma \eta$ are measured to be $\mathcal{B}(h_{c} \too \gamma \eta')=(1.52 \pm 0.27 \pm 0.29)\times10^{-3}$ and $\mathcal{B}(h_{c} \too \gamma \eta)=(4.7 \pm 1.5 \pm 1.4)\times10^{-4}$, respectively, where the first errors are statistical and the second are systematic uncertainties.
\end{abstract}


\maketitle
%%%%%%%%%%%%%%%%%%%%%%%%%%%%%%%%%%%%%%%%%%%%%%%%%%%%%%%%%%%%%%%%%%%%%%%%%%%%%%
%%%%%%%%%%%%%%%%%%%%%%%%%%%%%%%%%%%%%%%%%%%%%%%%%%%%%%%%%%%%%%%%%%%%%%%%%%%%%%

Charmonium, the bound state of a charmed quark and anticharmed quark ($c\bar{c}$), has played an important role for our understanding of quantum chromodynamics (QCD), which is the fundamental theory that describes the strong interactions between quarks and gluons. At low energies, QCD remains of high interest both experimentally and theoretically. All charmonium states below the open-charm $D\bar{D}$ threshold have been observed experimentally and can be well described by potential models~\cite{model}. However, knowledge is still sparse on the $P$-wave spin-singlet state, $h_c(^{1}P_{1})$. So far, only a few decay modes of $h_c$ have been observed, in particular, the radiative transition $h_c\too\gamma\eta_{c}$ (with a branching fraction $\mathcal{B} \approx 50\%$)~\cite{cleoc} and one hadronic decay $h_c\too2(\pi^{+}\pi^{-})\pi^{0}$ ($\mathcal{B} \approx 2\%$)~\cite{cleochadron}. Searches for the new $h_{c}$ decay modes, such as $h_{c} \too \gamma \eta(\eta')$, are useful for providing constraints to theoretical models in the charmonium region. The ratio of the branching fraction $\mathcal{B}(h_{c} \too \gamma \eta)$ over $\mathcal{B}(h_{c} \too \gamma \eta')$ can also be used to study the $\eta-\eta'$ mixing angle~\cite{mixing}, which is important to test SU(3)-flavor symmetries in QCD.

First evidence for the decay mode $h_{c} \too \gamma\eta_{c}$ was seen by the E835 experiment in $p\bar{p}$ collisions~\cite{e835} with a significance of about 3$\sigma$. This was subsequently confirmed by CLEO-c~\cite{cleoc} in the decay chain $\psi' \too \pi^{0}h_{c}, h_{c} \too \gamma\eta_{c}$, where $\psi'$ is shorthand for $\psi(3686)$. Recently, the BESIII experiment improved accuracy of the $h_{c}$ decay properties with $1.06 \times 10^{8}$ $\psi'$ events in $\psi' \too \pi^{0}h_{c}, h_{c} \too \gamma\eta_{c}$~\cite{bes1, bes2}. The spin-singlet state $h_{c}$ cannot be produced directly in $\EE$ collisions, but it can be produced through $\psi' \too \pi^{0}h_{c}$ with a production rate of the order of $10^{-3}$. Since the $h_{c}$ has negative $C$-parity, it very likely decays into a photon plus a pseudoscalar meson, such as $\eta'$ and $\eta$.

In this paper, we report the observation (evidence) of the $h_{c}$ radiative decay $h_{c} \too \gamma\eta'(\eta)$, where $h_{c}$ is produced in the decay $\psi' \too \pi^{0}h_{c}$. The $h_{c} \too \gamma\eta'$ is reconstructed by using $\eta' \too \pi^{+}\pi^{-}\eta$ with $\eta \too \gamma\gamma$ and $\eta' \too \gamma\pi^{+}\pi^{-}$. The $h_{c} \too \gamma\eta$ is reconstructed from decays $\eta \too \gamma\gamma$ and $\eta \too \pi^{+}\pi^{-}\pi^{0}$ with $\pi^{0} \too \gamma\gamma$. The analyses are based on a data sample of $4.48 \times 10^{8}$ $\psi'$ events collected with the BESIII detector~\cite{besiii} in 2009 and 2012. The number of $\psi'$ events is $(1.069 \pm 0.075)$ $\times 10^{8}$ for 2009 and $(3.411 \pm 0.021)$ $\times 10^{8}$ for 2012 from counting inclusive hadronic events~\cite{totalnumber}. A data sample of 44 pb$^{-1}$ integrated luminosity, taken at center-of-mass energy $\sqrt{s} = 3.65$ GeV, is used to estimate the background contribution from continuum processes.  Samples of Monte Carlo (MC) simulated events for the signal decay $\psi' \too \pi^{0}h_{c}, h_{c} \too \gamma\eta'(\eta)$ are generated using the HELAMP model in {\sc evtgen}~\cite{evtgen}. A Monte Carlo (MC) sample of generic $\psi'$ events (``inclusive MC'') is used for background studies. The $\psi'$ resonance is produced by the event generator {\sc kkmc}~\cite{kkmc}, and the decays are generated by {\sc evtgen}~\cite{evtgen} with known branching fractions~\cite{pdg}, while unmeasured decays are generated according to the {\sc lundcharm} model~\cite{lund}.

The BESIII detector has a geometrical acceptance of 93$\%$ of 4$\pi$. A small cell helium-based main drift chamber (MDC) provides momentum measurements of charged particles; in a 1 T magnetic field, the momentum resolution is 0.5$\%$ at 1 GeV/$c$. It also supplies an energy loss ($dE/dx$) measurement with a resolution better than 6$\%$ for electrons from Bhabha scattering. The electromagnetic calorimeter (EMC) measures photon energies with a resolution of 2.5$\%$ (5$\%$) at 1 GeV in the barrel (endcaps). The time-of-flight system (TOF) is composed of plastic scintillators with a time resolution of 80 ps (110 ps) in the barrel (endcap) and is used for charged particle identification.

Each charged track is required to have a point of closest approach to the beamline within 1 cm in the radial direction and within 10 cm from the interaction point (IP) along the beam direction. The polar angle of the tracks must be well contained within the fiducial volume of the MDC, $|\cos\theta|<0.93$ in the laboratory frame. Photons are reconstructed from isolated showers in the EMC that are at least $10^\circ$ away from the nearest charged track. The photon energy deposition is required to be at least 25 MeV in the barrel region of the EMC $(|\cos\theta|<0.8)$ or 50 MeV in the EMC endcaps $(0.86<|\cos\theta|<0.92)$. In order to suppress electronic noise and energy depositions that are unrelated to the event, the EMC time $t$ of the photon candidates must be in coincidence with collision events within the range $0 \leq t \leq 700$ ns. This criterion is applied only when there are charged particles in the final state.

For the decay chains $\psi' \too \pi^{0}h_{c}$, where $h_{c} \too \gamma\eta'(\eta' \too \pi^{+}\pi^{-}\eta)$ or $h_{c} \too \gamma\eta(\eta \too \pi^{+}\pi^{-}\pi^{0})$, both final states have five photons and a $\pi^{+}\pi^{-}$ pair. A vertex fit is performed on the two charged tracks to ensure that the tracks originate from the IP. In order to reduce background events and to improve the mass resolution, a $6C$-kinematic fit is performed imposing overall energy and momentum conservation and constraining the masses of the $\pi^{0}$ and $\eta$ mesons to their nominal values~\cite{pdg} in the $h_{c} \too \gamma\eta'(\eta' \too \pi^{+}\pi^{-}\eta)$ decay and the masses of two $\pi^{0}$'s to the nominal mass in the $h_{c} \too \gamma\eta(\eta \too \pi^{+}\pi^{-}\pi^{0})$ decay. We loop over all possible combinations of photons, and select the one with the least $\chi^{2}_{6C}$ of the kinematic fit. The $\chi^{2}_{6C}$ of a candidate event is required to be less than 120. For the $h_{c} \too \gamma\eta'(\eta' \too \gamma\pi^{+}\pi^{-})$ decay chain, the final state has four photons and a $\pi^{+}\pi^{-}$ pair. A vertex fit is applied on the two charged tracks and a $5C$-kinematic fit is performed imposing conservation of the initial four-momentum and constraining the mass of the $\pi^{0}$ meson to its nominal value. We loop over all possible combinations of photons, selecting the combination with the least $\chi^{2}_{5C}$ of the kinematic fit. The $\chi^{2}_{5C}$ of candidate events is required to be less than 50. Of the two photons, the one with the larger energy is selected as the radiative photon from $h_{c}$. For the $h_{c} \too \gamma\eta(\eta \too \gamma\gamma)$ analysis the final state has only five photons. A 6C-kinematic fit is performed to the total initial four-momentum of the colliding beams, while the masses of the $\pi^{0}$ and $\eta$ mesons are constrained to their nominal values. We loop over all possible combinations of photons and select the ones with the least $\chi^{2}_{6C}$ of the kinematic fit. In order to be able to use the $\eta$ sideband to verify signals, for the selected five photons a 5C-kinematic is performed constraining the four-momentum of the final state to the total initial four-momentum of the $\EE$ beams and the mass of the $\pi^{0}$ meson to its nominal value. The $\chi^{2}_{5C}$ of candidate events is required to be less than 35. All the selection criteria have been optimized by maximizing the figure of merit $S/\sqrt{S+B}$, where $S (B)$ is the number of signal (background) events in the signal region.

\begin{figure}[htbp]
\begin{center}
\begin{overpic}[width=0.22\textwidth]{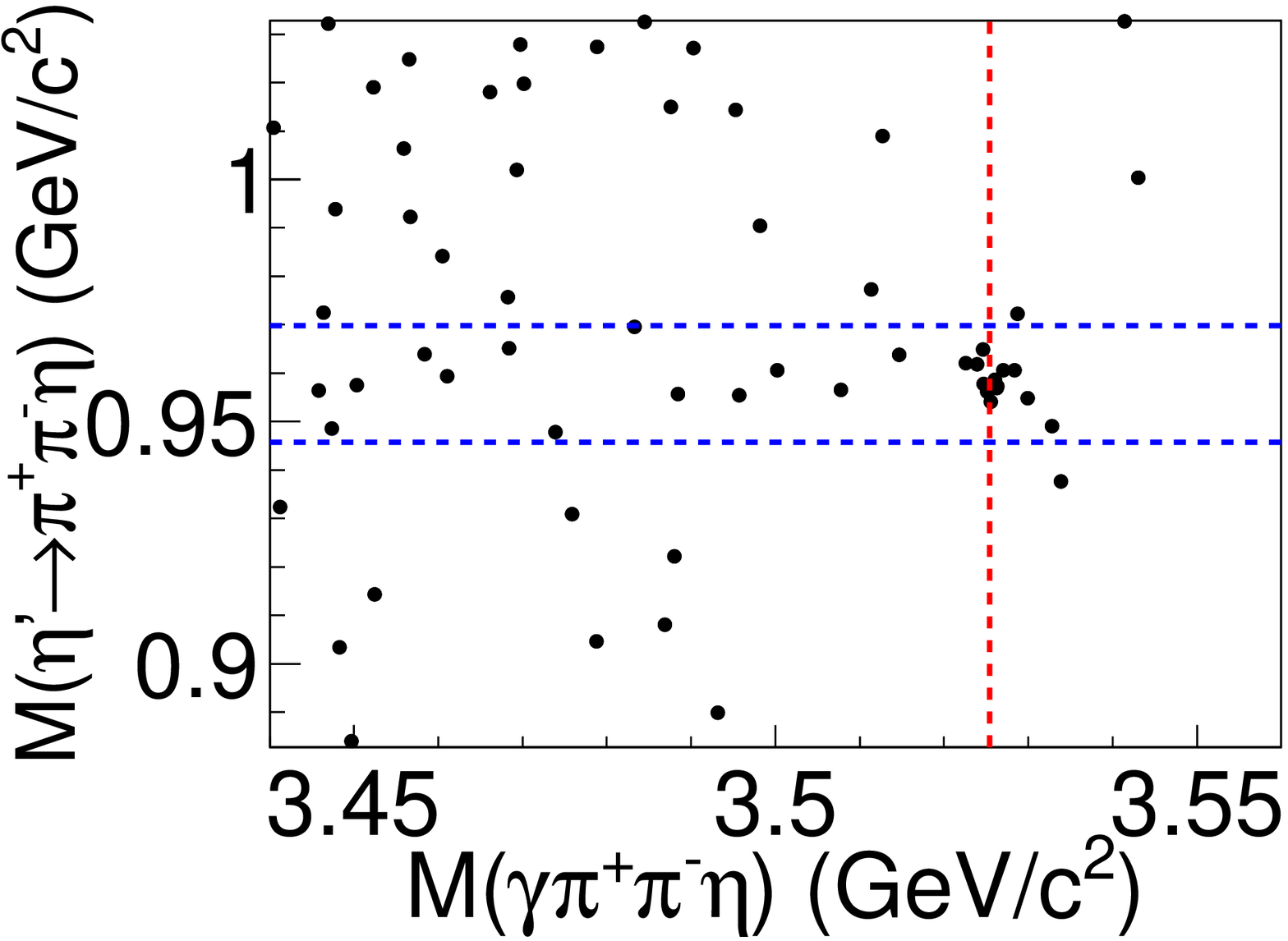}
\put(91,65){(a)}
\end{overpic}
\begin{overpic}[width=0.22\textwidth]{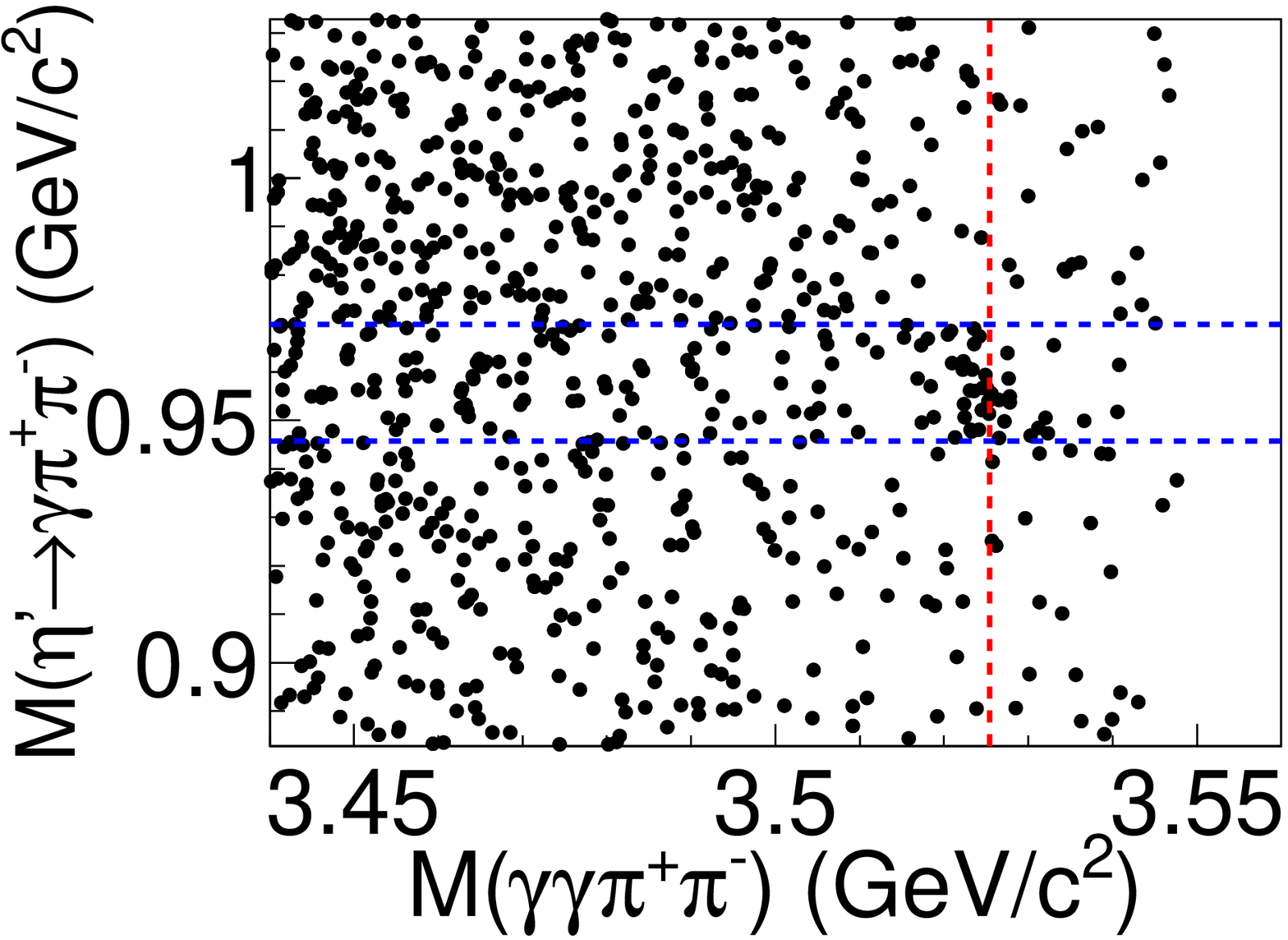}
\put(91,65){(b)}
\end{overpic}
\begin{overpic}[width=0.22\textwidth]{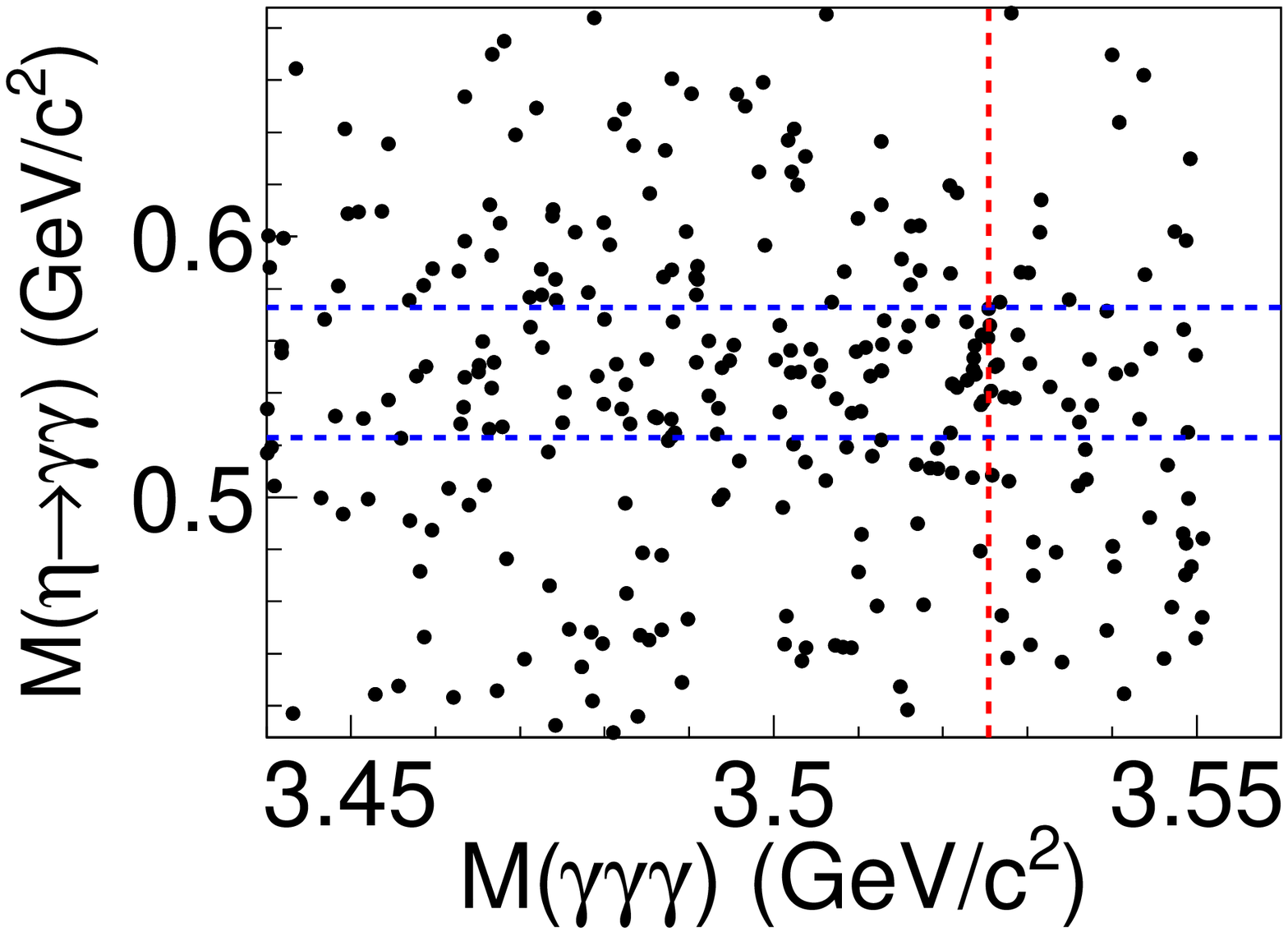}
\put(91,65){(c)}
\end{overpic}
\begin{overpic}[width=0.22\textwidth]{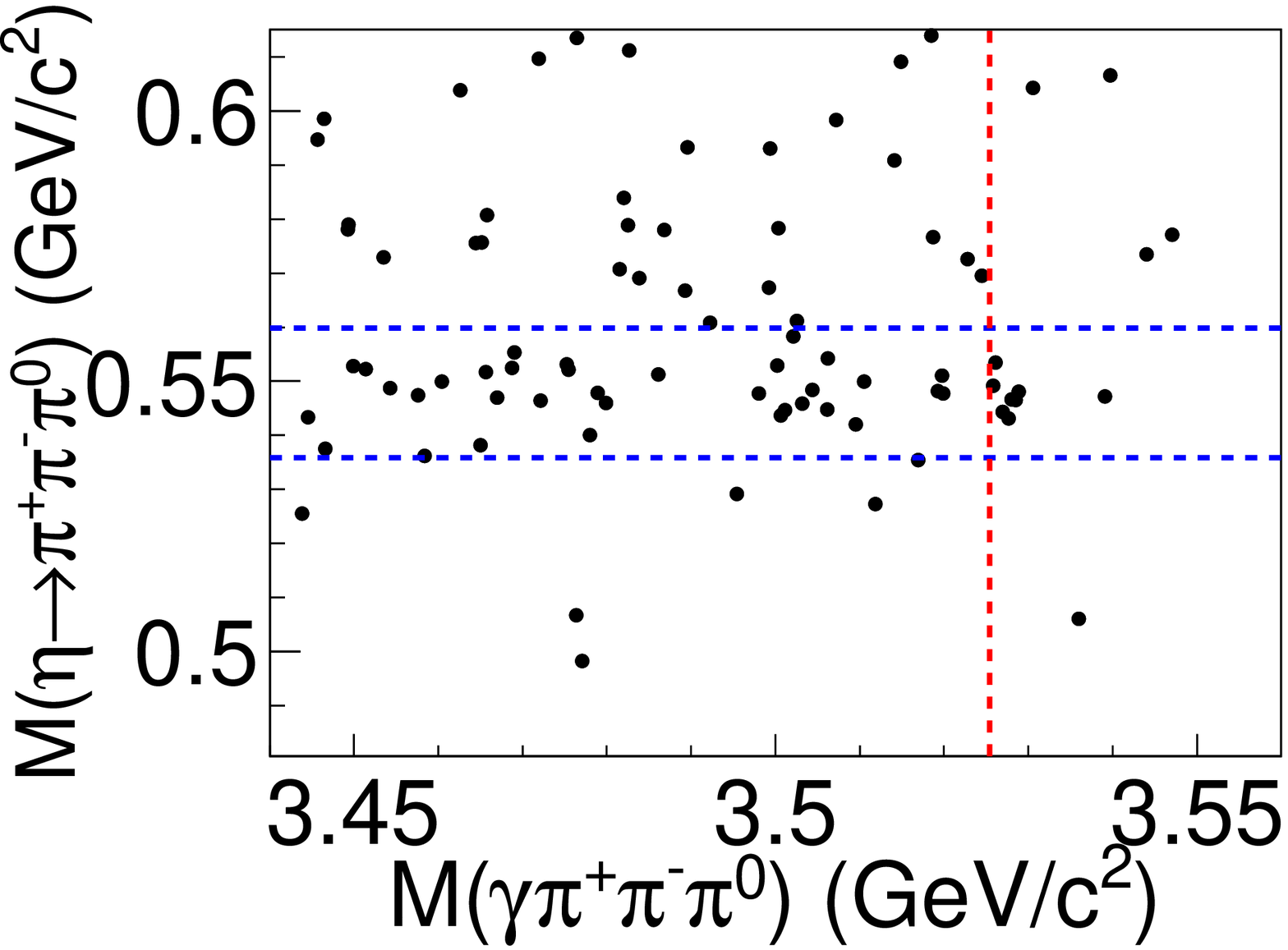}
\put(91,65){(d)}
\end{overpic}
\caption{Scatter plots of the selected events from the $\psi'$ data set. (a) $M(\eta' \too \pi^{+}\pi^{-}\eta)$ versus $M(\gamma\pi^{+}\pi^{-}\eta)$ for $h_{c} \too \gamma\eta'(\eta' \too \pi^{+}\pi^{-}\eta)$.
(b) $M(\eta' \too \gamma\pi^{+}\pi^{-})$ versus $M(\gamma\gamma\pi^{+}\pi^{-})$ for $h_{c} \too \gamma\eta'(\eta' \too \gamma\pi^{+}\pi^{-})$.
(c) $M(\eta \too \gamma\gamma)$ versus $M(\gamma\gamma\gamma)$ for $h_{c} \too \gamma\eta(\eta \too \gamma\gamma)$.
(d) $M(\eta \too \pi^{+}\pi^{-}\pi^{0})$ versus $M(\gamma\pi^{+}\pi^{-}\pi^{0})$ for $h_{c} \too \gamma\eta(\eta \too \pi^{+}\pi^{-}\pi^{0})$.
The blue dashed lines mark the signal region of $\eta'$ $(\eta)$ and the red dashed lines mark the nominal $h_{c}$ mass.
}
\label{fig:scatter}
\end{center}
\end{figure}

With the above selection requirements applied, scatter plots for the decay $h_{c} \too \gamma\eta'$ are shown in Fig.~\ref{fig:scatter} as plot (a) for $\eta' \too \pi^{+}\pi^{-}\eta$ and plot (b) for $\eta' \too \gamma\pi^{+}\pi^{-}$. Clear enhancements are seen in the $\eta'$ and $h_{c}$ signal regions. The $\eta'$ signal region is defined as [$M_{\eta'}-$12, $M_{\eta'}$+12] MeV/$c^{2}$. The regions [$M_{\eta'}-$60, $M_{\eta'}-$36] MeV/$c^{2}$ and [$M_{\eta'}$+36, $M_{\eta'}$+60] MeV/$c^{2}$ are taken as the $\eta'$ sidebands, which are twice as wide as the signal region, where $M_{\eta'}$ is the nominal mass of the $\eta'$~\cite{pdg}. The scatter plots for the decay $h_{c} \too \gamma\eta$ are shown in plot (c) for $\eta \too \gamma\gamma$ and plot (d) for $\eta \too \pi^{+}\pi^{-}\pi^{0}$. An accumulation of events can be seen in the $\eta$ and $h_{c}$ signal regions. For the $\eta \too \gamma\gamma$ decay mode, where the mass resolution is about 8 MeV/$c^{2}$, the $\eta$ signal region is defined as [$M_{\eta}-$25, $M_{\eta}$+25] MeV/$c^{2}$, and the regions [$M_{\eta}-$100, $M_{\eta}-$50] MeV/$c^{2}$ and [$M_{\eta}$+50, $M_{\eta}$+100] MeV/$c^{2}$ are taken as the $\eta$ sidebands. For the $\eta \too \pi^{+}\pi^{-}\pi^{0}$ decay mode, where the mass resolution is about 3 MeV/$c^{2}$, the $\eta$ signal region is defined as [$M_{\eta}-$12, $M_{\eta}$+12] MeV/$c^{2}$, and the regions [$M_{\eta}-$48, $M_{\eta}-$24] MeV/$c^{2}$ and [$M_{\eta}$+24, $M_{\eta}$+48] MeV/$c^{2}$ are taken as the $\eta$ sideband, where $M_{\eta}$ is the nominal mass of $\eta$.

Possible background contributions are studied with the $\psi'$ inclusive MC sample and with the continuum data set collected at a center-of-mass energy of $\sqrt{s}=3.65$ GeV. From the latter, none of the continuum events survive the event selection requirement. The study with the $\psi'$ inclusive MC sample shows that the main background processes are $\pi^{0}\pi^{0}J/\psi(\gamma\eta')$ and $\omega(\gamma\pi^{0})\eta'$ for the $\eta' \too \pi^{+}\pi^{-}\eta$ decay mode; $\omega(\gamma\pi^{0})\eta'$ and $\gamma\chi_{c0}(\rho^{+}\rho^{-})$ for the $\eta' \too \gamma\pi^{+}\pi^{-}$ decay mode; and $\gamma\chi_{c2}(\eta\eta)$ for the $\eta \too \gamma\gamma$ and $\eta \too \pi^{+}\pi^{-}\pi^{0}$ decay modes. None of the background channels shows a peaking behaviour in the signal region, and their overall contribution is found to be smooth.

\begin{figure}[htbp]
\begin{center}
\begin{overpic}[width=0.22\textwidth]{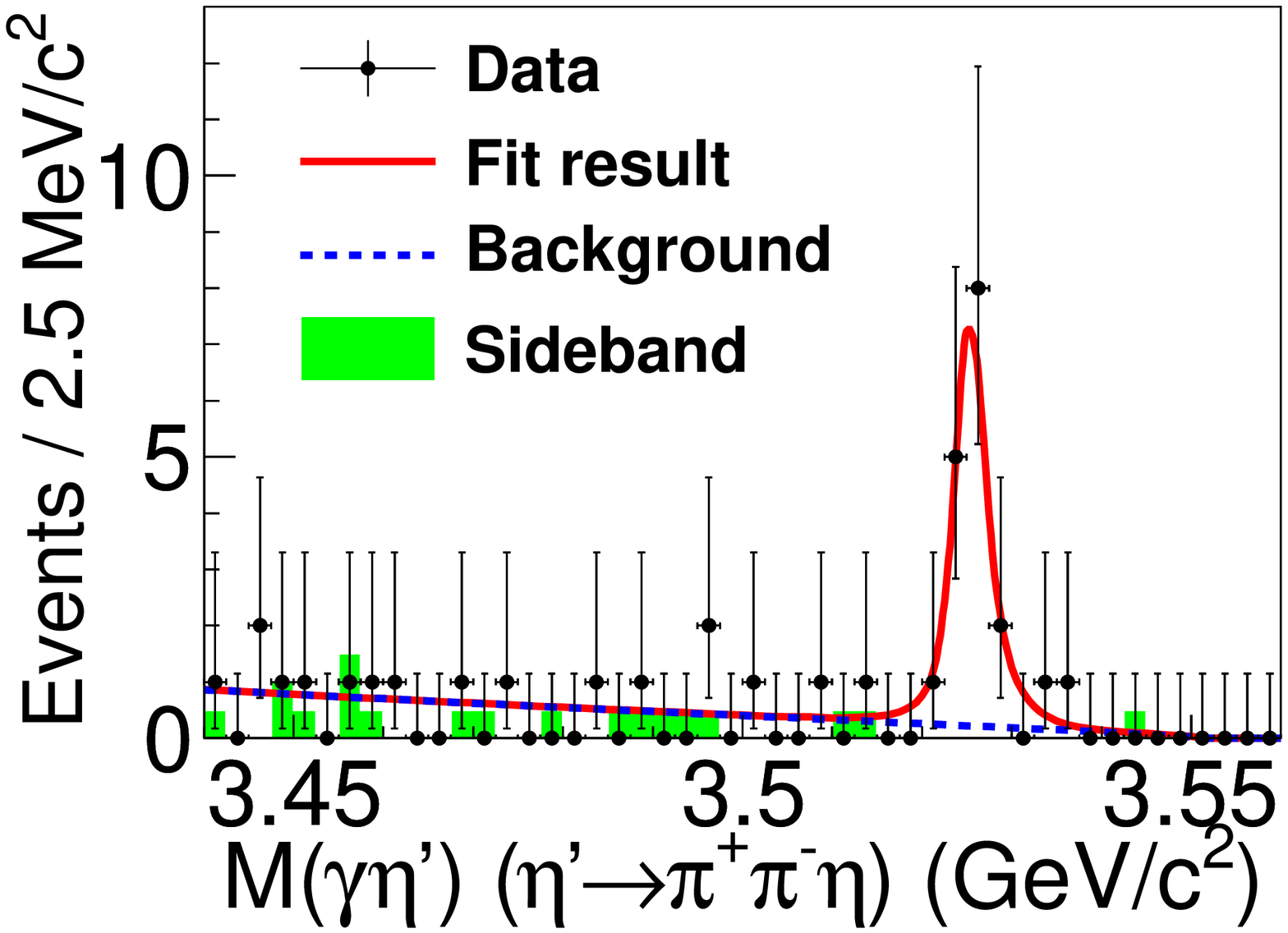}
\put(91,65){(a)}
\end{overpic}
\begin{overpic}[width=0.22\textwidth]{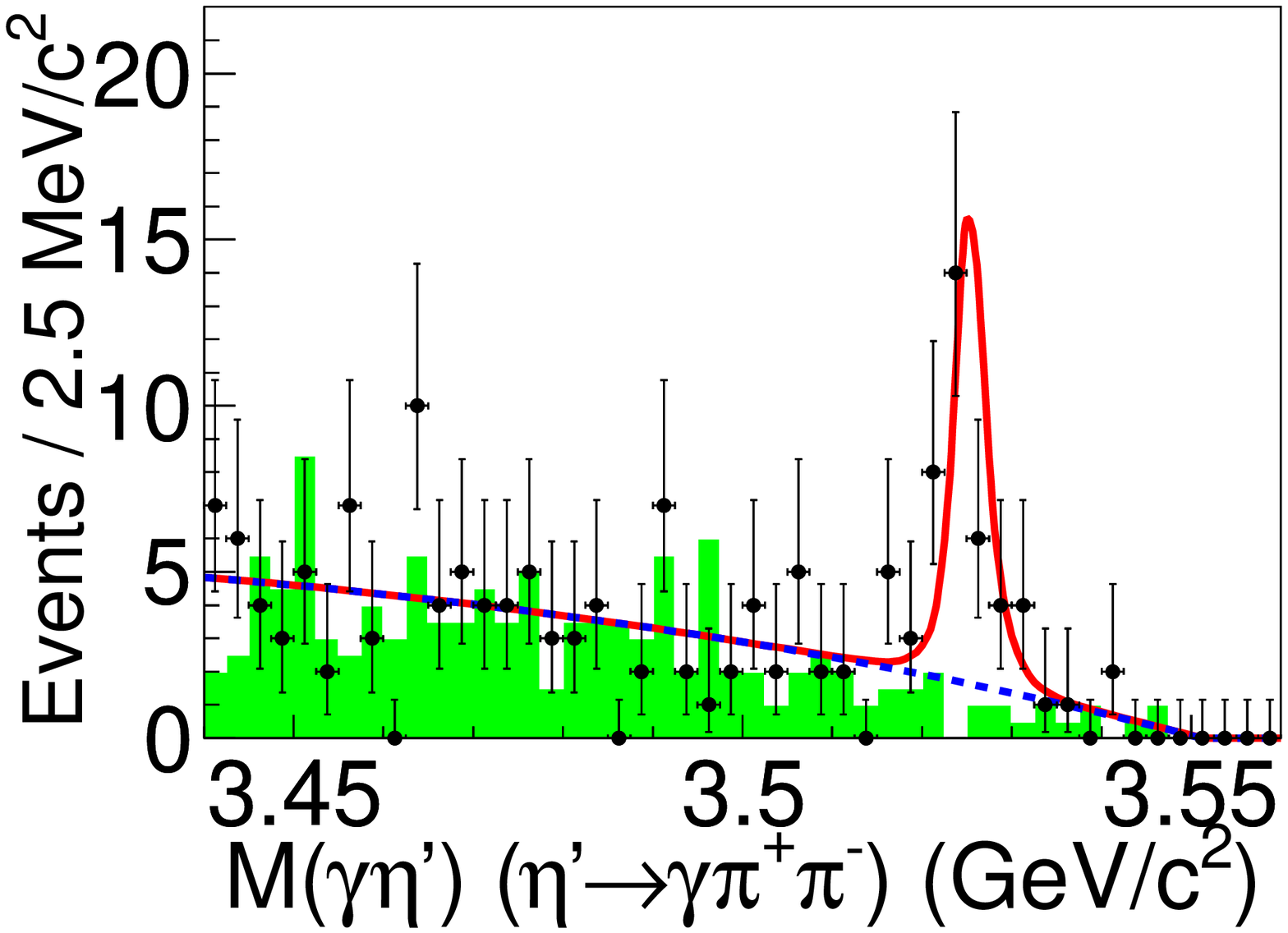}
\put(91,65){(b)}
\end{overpic}
\begin{overpic}[width=0.22\textwidth]{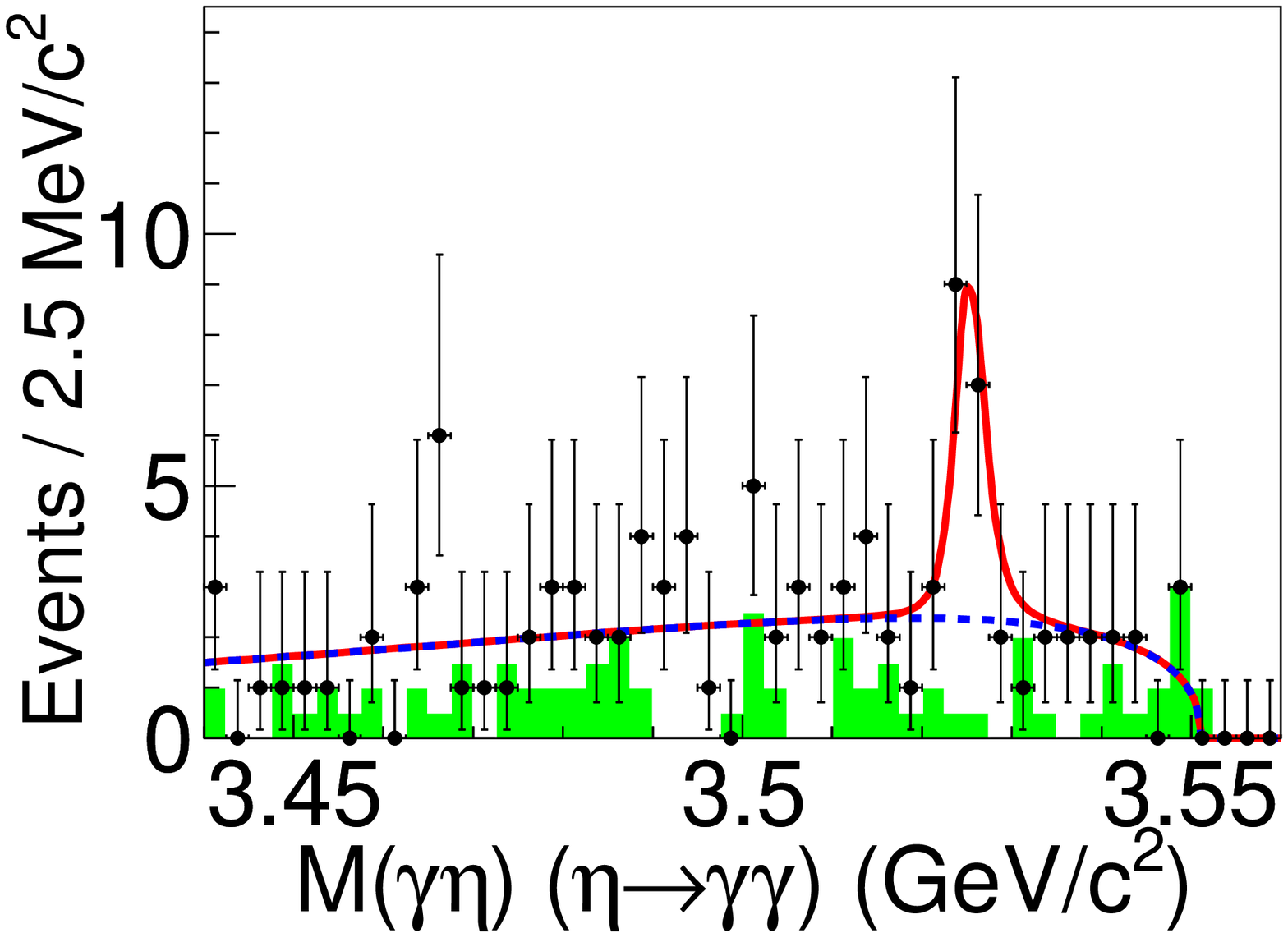}
\put(91,65){(c)}
\end{overpic}
\begin{overpic}[width=0.22\textwidth]{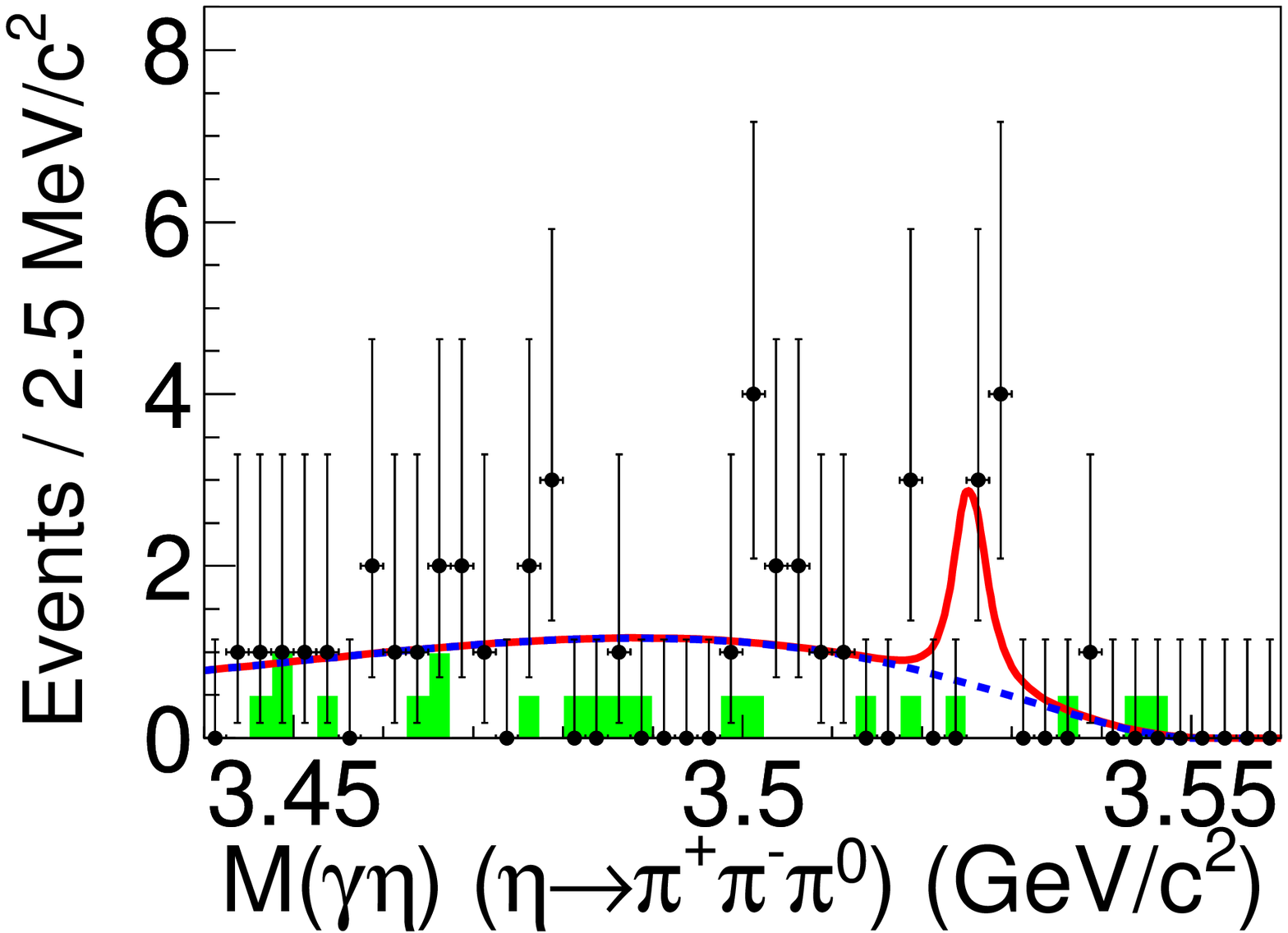}
\put(91,65){(d)}
\end{overpic}
\caption{Results of the simultaneous fits to the two invariant mass distributions of (top) $M(\gamma \eta')$ and (below) $M(\gamma \eta)$ for data. (a) $M(\gamma\eta')$ distribution for $h_{c} \too \gamma\eta'(\eta' \too \pi^{+}\pi^{-}\eta)$. (b) $M(\gamma\eta')$ distribution for $h_{c} \too \gamma\eta'(\eta' \too \gamma\pi^{+}\pi^{-})$. (c) $M(\gamma\eta)$ distribution for $h_{c} \too \gamma\eta(\eta \too \gamma\gamma)$. (d) $M(\gamma\eta)$ distribution for $h_{c} \too \gamma\eta(\eta \too \pi^{+}\pi^{-}\pi^{0})$. The red solid curves are the fit results, the blue dashed curves are the background distributions, and the green hatched histograms are events from the $\eta' (\eta)$ sidebands.}
\label{fig:fitresult}
\end{center}
\end{figure}

Figure~\ref{fig:fitresult} shows the distributions of the invariant masses $M(\gamma \eta')$ and $M(\gamma \eta)$ for the selected events. Signals of the $h_{c}$ meson are observed. In order to extract the signal yield a simultaneous maximum likelihood fit is performed on $\eta' \too \pi^{+}\pi^{-}\eta$ and $\eta' \too \gamma\pi^{+}\pi^{-}$ events for the $h_{c} \too \gamma\eta'$ decay, and on $\eta \too \gamma\gamma$ and $\eta \too \pi^{+}\pi^{-}\pi^{0}$ events for the $h_{c} \too \gamma\eta$ decay, respectively. The signal shape is modelled using signal MC events. The background is described with the ARGUS function~\cite{argus}:
\begin{equation}
  m\cdot(1-(m/m_{0})^{2})^{p}\cdot{\exp(k(1-(m/m_{0})^{2}))}\cdot\theta(m<m_{0}) ,
\end{equation}
where $p$ and $k$ are free parameters in the fit, and $m_{0}$ is fixed at $\sqrt{s}-M_{\pi^{0}}$, $M_{\pi^{0}}$ is the nominal $\pi^{0}$ mass. In the fit, the ratio of the number of $\eta' \too \pi^{+}\pi^{-}\eta$ signal events to the number of $\eta' \too \gamma\pi^{+}\pi^{-}$ signal events is fixed at $\frac{\mathcal{B}(\eta' \too \pi^{+}\pi^{-}\eta)\cdot{\mathcal{B}(\eta \too \gamma\gamma)}\cdot{\epsilon_{\eta' \too \pi^{+}\pi^{-}\eta}}}{\mathcal{B}(\eta' \too \gamma\pi^{+}\pi^{-})\cdot{\epsilon_{\eta' \too \gamma\pi^{+}\pi^{-}}}}=0.515\pm0.013$, where $\epsilon_{\eta' \too\pi^{+}\pi^{-}\eta}$ and $\epsilon_{\eta' \too \gamma\pi^{+}\pi^{-}}$ are the global efficiencies for the reconstruction of events of the channel $\psi' \too \pi^{0}h_{c}, h_{c} \too \gamma\eta', \eta' \too \pi^{+}\pi^{-}\eta$ and $\psi' \too \pi^{0}h_{c}, h_{c} \too \gamma\eta', \eta' \too \gamma\pi^{+}\pi^{-}$ decay modes, respectively, determined from MC simulations. The $\eta^{(')}$ branching fractions are taken from the Particle Data Group (PDG)~\cite{pdg}. Similarly the ratio of the number of $\eta \too \gamma\gamma$ events to the number of $\eta \too \pi^{+}\pi^{-}\pi^{0}$ is fixed at $\frac{\mathcal{B}(\eta \too \gamma\gamma)\cdot{\epsilon_{\eta \too \gamma\gamma}}}{\mathcal{B}(\eta \too \pi^{+}\pi^{-}\pi^{0})\cdot{\mathcal{B}(\pi^{0} \too \gamma\gamma)}\cdot{\epsilon_{\eta \too \pi^{+}\pi^{-}\pi^{0}}}}=2.597\pm0.006$. The fit results are shown as the solid curves in Fig.~\ref{fig:fitresult}. For the $h_{c} \too \gamma\eta'$ decay, the total $h_{c}$ signal yield is $N_{h_{c}\too\gamma\eta'}=44.3\pm7.8$. The statistical significance of the $h_{c}$ signal is $8.4\sigma$ as found by comparing the likelihood values ($\Delta(\ln\mathcal{L})=35.4$) for the fits with or without $h_{c}$ signal and taking the change of the number of degrees-of-freedom ($\Delta \text{ndf}=1$) into account. The goodness of the fit is $\chi^{2}/\text{ndf}=12.9/14=0.9$. For the $h_{c} \too \gamma\eta$ decay, the signal yield is $N_{h_{c}\too\gamma\eta}=18.1\pm5.8$ with a statistical significance of $4.0\sigma$($\Delta(\ln\mathcal{L})=8.0$, $\Delta \text{ndf}=1$), and the goodness of the fit is $\chi^{2}/\text{ndf}=14.0/10=1.4$.

The branching fractions $\mathcal{B}(h_{c} \too \gamma \eta')$ and $\mathcal{B}(h_{c} \too \gamma \eta)$ are calculated using the following formulae:\begin{footnotesize}
\begin{equation}
    \mathcal{B}(h_{c} \too \gamma \eta^{(')}) = \frac{N_{h_{c}\too\gamma\eta^{(')}}}{N_{\psi'}\cdot{\mathcal{B}(\psi' \too \pi^{0}h_{c})}\cdot{\mathcal{B}(\pi^{0} \too \gamma\gamma)}\cdot W_{\eta^{(')}}} ,
\end{equation}
\end{footnotesize}
where $W_{\eta'}$ is the sum of ${\mathcal{B}(\eta' \too \pi^{+}\pi^{-}\eta)}\cdot{\mathcal{B}(\eta \too \gamma\gamma)}\cdot{\epsilon}_{\eta' \too \pi^{+}\pi^{-}\eta}$ and ${\mathcal{B}(\eta' \too \gamma\pi^{+}\pi^{-})}\cdot{\epsilon}_{\eta' \too \gamma\pi^{+}\pi^{-}}$, $W_{\eta}$ is the sum of $\mathcal{B}(\eta \too \gamma\gamma)\cdot{\epsilon}_{\eta\too\gamma\gamma}$ and ${\mathcal{B}(\eta \too \pi^{+}\pi^{-}\pi^{0})}\cdot{\mathcal{B}(\pi^{0} \too \gamma\gamma)}\cdot{\epsilon}_{\eta\too\pi^{+}\pi^{-}\pi^{0}}$, $N_{h_{c}\too\gamma\eta'}$ $(N_{h_{c}\too\gamma\eta})$ is the observed number of $h_{c}\too\gamma\eta'$ $(h_{c}\too\gamma\eta)$ signal events, and $N_{\psi'}$ is the observed number of $\psi'$ events in the data set. The corresponding branching fractions of $h_{c} \too \gamma \eta'$ and $h_{c} \too \gamma \eta$ are measured to be $(1.52 \pm 0.27)\times10^{-3}$ and $(4.7 \pm 1.5)\times10^{-4}$, where the errors are statistical. The results for $h_{c} \too \gamma\eta'(\eta)$ are listed in Table~\ref{tab:branching}.

\begin{table*}[htbp]
\begin{center}
\caption{ Results on $h_{c} \too \gamma\eta'(\eta)$. The table shows the decay mode, total number of events $N_{h_{c} \too \gamma\eta'(\eta)}$, the daughter branching fraction $W_{\eta'}={\mathcal{B}(\eta' \too \pi^{+}\pi^{-}\eta)}\cdot{\mathcal{B}(\eta \too \gamma\gamma)}\cdot{\epsilon}_{\eta' \too \pi^{+}\pi^{-}\eta}+{\mathcal{B}(\eta' \too \gamma\pi^{+}\pi^{-})}\cdot{\epsilon}_{\eta' \too \gamma\pi^{+}\pi^{-}}$, $W_{\eta}=\mathcal{B}(\eta \too \gamma\gamma)\cdot{\epsilon}_{\eta\too\gamma\gamma}+{\mathcal{B}(\eta \too \pi^{+}\pi^{-}\pi^{0})}\cdot{\mathcal{B}(\pi^{0} \too \gamma\gamma)}\cdot{\epsilon}_{\eta\too\pi^{+}\pi^{-}\pi^{0}}$, measured branching fractions $\mathcal{B}(h_{c} \too \gamma \eta'(\eta))$, statistical significance, and the ratio of the branching fractions $\mathcal{B}(h_{c} \too \gamma \eta)$ over $\mathcal{B}(h_{c} \too \gamma \eta')$. }
\label{tab:branching}
\begin{tabular}{cccccc}
  \hline
  \hline
  Mode & $N_{h_{c} \too \gamma\eta'(\eta)}$ & $W_{\eta'(\eta)}$($\times10^{-2}$) & $\mathcal{B}(h_{c} \too \gamma \eta'(\eta))$ & Significance &  $\frac{\mathcal{B}(h_{c} \too \gamma \eta)}{\mathcal{B}(h_{c} \too \gamma \eta')}(\%)$ \\
  \hline
  $h_{c} \too \gamma\eta'$  & 44.3 $\pm$ 7.8(stat.)      & 7.67 $\pm$ 0.38(sys.)       & (1.52 $\pm$ 0.27(stat.) $\pm$ 0.29(sys.))$\times10^{-3}$      & 8.4$\sigma$  & \multirow{2}*{$30.7 \pm 11.3$(stat.) $\pm 8.7$(sys.)} \\
  $h_{c} \too \gamma\eta $  & 18.1 $\pm$ 5.8(stat.)      & 10.22 $\pm$ 0.55(sys.)      & (4.7 $\pm$ 1.5(stat.) $\pm$ 1.4(sys.))$\times10^{-4}$      & 4.0$\sigma$ &  \\
  \hline
  \hline
\end{tabular}
\end{center}
\end{table*}

Systematic uncertainties in the branching fractions measurement for $h_{c} \too \gamma\eta'(\eta)$ originate mainly from the data/MC difference in the tracking efficiency, photon detection, $\pi^{0}/\eta$ reconstruction, and the kinematic fit, as well as from MC statistics, the branching fractions taken from world averages~\cite{pdg}, the total number of $\psi'$ events in the data set, the fit range, the signal and background shapes.

The difference between data and MC in tracking efficiency for each charged track is estimated to be $1\%$~\cite{trackeff}, and so a $2\%$ systematic uncertainty is given to all channels with charged tracks. The uncertainty due to photon detection efficiency is determined by using events of the control sample $J/\psi \too \rho^{0}\pi^{0}$ and found to be $1.0\%$ per photon~\cite{rhopi}.

\begin{table*}[htbp]
\caption{Summary of systematic uncertainties (in units of $\%$).}
\label{tab:sumerror}
%\begin{footnotesize}
\begin{tabular}{ c | c  c | c  c }
  \hline
  \hline
  Source & $\eta' \too \pi^{+}\pi^{-}\eta$ & $\eta' \too \gamma\pi^{+}\pi^{-}$ & $\eta \too \gamma\gamma$ & $\eta \too \pi^{+}\pi^{-}\pi^{0}$  \\
  \hline
  Tracking & 2.0 & 2.0 & - & 2.0 \\

  Photon & 5.0 & 4.0 & 5.0 & 5.0 \\

  $\pi^{0}$ and $\eta$ reconstruction & 2.0 & 1.0 & 2.0 & 2.0 \\

  Kinematic fit & 1.0 & 1.5 & 1.1 & 1.0\\

  MC statistics & 0.3 & 0.3 & 0.3 & 0.3\\

  $\mathcal{B}_{\eta',\eta,\pi^{0}}$ & 1.7 & 1.7 & 0.5 & 1.2 \\

  Number of $\psi'$ & \multicolumn{2}{c|}{0.7} & \multicolumn{2}{c}{0.7} \\

  Fit range & \multicolumn{2}{c|}{1.1} & \multicolumn{2}{c}{7.2} \\

  Signal shape & \multicolumn{2}{c|}{3.8} & \multicolumn{2}{c}{3.9} \\

  Background shape & \multicolumn{2}{c|}{9.7} & \multicolumn{2}{c}{24.9} \\

  $\mathcal{B}(\psi' \too \pi^{0}h_{c})\cdot{\mathcal{B}(\pi^{0} \too \gamma\gamma)}$ & \multicolumn{2}{c|}{15.1} & \multicolumn{2}{c}{15.1} \\

  Sum & \multicolumn{2}{c|}{19.1} & \multicolumn{2}{c}{30.7} \\
  \hline
  \hline
\end{tabular}
%\end{footnotesize}
\end{table*}

The uncertainty due to $\pi^{0}$ reconstruction is determined by using a high purity control sample of $J/\psi \too \pi^{0}p\bar{p}$ decays~\cite{piefficiency}. The efficiency for the $\pi^{0}$ reconstruction is obtained from the $\pi^{0}$ yields determined from the $\pi^{0}$ mass spectrum with or without the $\pi^{0}$ selection requirements. The difference of the $\pi^{0}$ reconstruction efficiency between data and MC simulation is found to be $1\%$ per $\pi^{0}$. The uncertainty of the $\eta$ reconstruction from $\gamma\gamma$ final states is $1\%$ per $\eta$, which is determined from a high purity control sample of $J/\psi \too \eta p\bar{p}$ in a similar way~\cite{piefficiency}.

For the uncertainty caused by the kinematic fit to the charged decay modes, we correct the track helix parameters in the MC so that the MC can better describe the momentum spectra of the data. In the analysis, we use the efficiency after the helix correction for the nominal results. The correction factors for pions are obtained by using the control sample $\psi' \too K^{+}K^{-}\pi^{+}\pi^{-}$~\cite{corrfactors}. The difference in the global efficiency between MC simulations performed before and after the correction is taken as the systematic uncertainty due to the kinematic fit. For the mode with only neutral particles in the final state the systematic uncertainty of the kinematic fit was studied with the non-resonant decay channel $J/\psi \too \gamma \eta \pi^{0}$, as they have the same final state.

The statistical uncertainty of the global efficiency determined from MC simulations is 0.3$\%$. The systematic uncertainties of the branching fractions are taken from the PDG~\cite{pdg}. The total number of $\psi'$ decay events is estimated by measuring inclusive hadronic events, as described in Ref.~\cite{totalnumber}.  The uncertainty of the total number of $\psi'$ events is estimated to be $0.7\%$.

The uncertainty due to the fit procedure includes the fit range, signal shape and background shape. The uncertainty due to the fit range is obtained by varying the limits of the fit range by $\pm$0.01 GeV/$c^{2}$, and the change in the final result is taken as the uncertainty. The uncertainty due to the signal shape is derived from the difference in the mass resolution between data and MC simulation, and from the errors of the $h_{c}$ resonance parameters. To study the differences in the mass resolution between data and MC simulation the $J/\psi$ distribution of the reaction $\psi' \too \eta J/\psi$($J/\psi \too \gamma\eta'$) is fitted with the MC shape of the $J/\psi$ convoluted with a Gaussian function. The parameters (mean $m$ and sigma $\sigma$) of the Gaussian function are determined to be $m=0.1\pm0.1$MeV, $\sigma = 0.6\pm0.3$MeV for $\eta' \too \pi^{+}\pi^{-}\eta$, and $m=0.0\pm0.2$MeV, $\sigma = 0.1\pm0.4$MeV for $\eta' \too \gamma\pi^{+}\pi^{-}$, so the difference between data and MC simulation is small. To be conservative, we construct Gaussian smearing functions with the above measured mean and sigma varied by $\pm1\sigma$, and convolve the MC-determined $h_{c}$ shape with them and refit the data. We take the largest difference as the systematic uncertainty. To consider the uncertainties of the $h_{c}$ resonance parameters, the MC-determined shape convolved by a Gaussian with the mean and sigma given by the errors of the $h_{c}$ parameters~\cite{pdg}, is used as the signal shape for a refit of the data, and the difference is assigned as the systematic uncertainty. These two systematic uncertainties are added in quadrature, assuming they are independent, to obtain the systematic uncertainty on the signal shape. The uncertainty caused by the background shape is estimated by changing the background shape from an ARGUS function to a linear function. The difference between the two methods is taken as the systematic uncertainty on the background shape.

Table~\ref{tab:sumerror} summarizes all the systematic uncertainties of the different decay modes. The overall systematic errors are obtained by adding all systematic uncertainties in quadrature by assuming they are independent.

In summary, using the data sample of 4.48 $\times 10^{8}$ $\psi'$ events collected with the BESIII detector operating at the BEPCII storage ring, the radiative decay process $h_{c} \too \gamma \eta'$ is observed with a statistical significance of $8.4 \sigma$ for the first time, and we have evidence for the process $h_{c} \too \gamma \eta$ with a statistical significance of $4.0 \sigma$. The corresponding branching fractions of $h_{c} \too \gamma \eta'$ and $h_{c} \too \gamma \eta$ are measured to be $(1.52 \pm 0.27 \pm 0.29)\times10^{-3}$ and $(4.7 \pm 1.5 \pm 1.4)\times10^{-4}$, respectively, where the first errors are statistical and the second are systematic. The ratio of the branching fraction $\mathcal{B}(h_{c} \too \gamma \eta)$ over $\mathcal{B}(h_{c} \too \gamma \eta')$ is $R_{h_{c}}=\frac{\mathcal{B}(h_{c} \too \gamma \eta)}{\mathcal{B}(h_{c} \too \gamma \eta')}=(30.7 \pm 11.3$(stat.) $\pm 8.7$(sys.))$\%$, where the common systematic errors between $\mathcal{B}(h_{c} \too \gamma \eta)$ and $\mathcal{B}(h_{c} \too \gamma \eta')$ cancel out. Although the uncertainty is large, the $\eta-\eta'$ mixing angle can be extracted from $R_{h_{c}}$ to test SU(3)-flavor symmetries in QCD~\cite{mixing}, following the methods used for equivalent decays of the $J/\psi$ and $\psi'$ mesons~\cite{piefficiency,psidecay1,psidecay2}.
%%%%%%%%%%%%%%%%%%%%%%%%%%%%%%%%%%%%%%
%%%%%%%%%%%%%%%%%%%%%%%%%%%%%%%%%%%%%%

The BESIII collaboration thanks the staff of BEPCII and the IHEP computing center for their strong support. This work is supported in part by National Key Basic Research Program of China under Contract No. 2015CB856700; National Natural Science Foundation of China (NSFC) under Contracts Nos. 11125525, 11235011, 11322544, 11335008, 11425524, 11521505, Y61137005C; the Chinese Academy of Sciences (CAS) Large-Scale Scientific Facility Program; the CAS Center for Excellence in Particle Physics (CCEPP); the Collaborative Innovation Center for Particles and Interactions (CICPI); Joint Large-Scale Scientific Facility Funds of the NSFC and CAS under Contracts Nos. 11179007, U1232201, U1332201; CAS under Contracts Nos. KJCX2-YW-N29, KJCX2-YW-N45; 100 Talents Program of CAS; National 1000 Talents Program of China; INPAC and Shanghai Key Laboratory for Particle Physics and Cosmology; German Research Foundation DFG under Contract No. Collaborative Research Center CRC-1044; Istituto Nazionale di Fisica Nucleare, Italy; Koninklijke Nederlandse Akademie van Wetenschappen (KNAW) under Contract No. 530-4CDP03; Ministry of Development of Turkey under Contract No. DPT2006K-120470; Russian Foundation for Basic Research under Contract No. 14-07-91152; The Swedish Resarch Council; U.S. Department of Energy under Contracts Nos. DE-FG02-05ER41374, DE-SC-0010504, DE-SC0012069, DESC0010118; U.S. National Science Foundation; University of Groningen (RuG) and the Helmholtzzentrum fuer Schwerionenforschung GmbH (GSI), Darmstadt; WCU Program of National Research Foundation of Korea under Contract No. R32-2008-000-10155-0.

\end{document}